# Ultrabright room-temperature single-photon emission from nanodiamond nitrogen-vacancy centers with sub-nanosecond excited-state lifetime

S. Bogdanov[1,2], M. Shalaginov[1,2], A. Lagutchev[1,2], C.-C. Chiang[1,2], D. Shah[1,2], A.S. Baburin[3,4], I.A. Ryzhikov[3,5], I.A. Rodionov[3,4], A. Boltasseva[1,2] and V.M. Shalaev[1,2]

[1]*School of Electrical & Computer Engineering and Birck Nanotechnology Center, Purdue University, West Lafayette, IN 47907, USA*
[2]*Purdue Quantum Center, Purdue University, West Lafayette, IN 47907, USA*
[3]*FMNS REC, Bauman Moscow State Technical University, Moscow, 105005, Russia*
[4]*Dukhov Research Institute of Automatics, Moscow, 127055, Russia*
[5]*Institute for Theoretical and Applied Electromagnetics RAS, Moscow, 125412, Russia*

Ultrafast emission rates obtained from quantum emitters coupled to plasmonic nanoantennas have recently opened fundamentally new possibilities in quantum information and sensing applications. Plasmonic nanoantennas greatly improve the brightness of quantum emitters by dramatically shortening their fluorescence lifetimes. Gap plasmonic nanocavities that support strongly confined modes are of particular interest for such applications[1]. We demonstrate single-photon emission from nitrogen-vacancy (NV) centers in nanodiamonds coupled to nanosized gap plasmonic cavities with internal mode volumes about $10^4$ times smaller than the cubic vacuum wavelength. The resulting structures features sub-nanosecond NV excited-state lifetimes and detected photon rates up to 50 million counts per second. Analysis of the fluorescence saturation allows the extraction of the multi-order excitation rate enhancement provided by the nanoantenna. Efficiency analysis shows that the NV center is producing up to 0.25 billion photons per second in the far-field.

Characterizing single photon sources[2], such as single molecules, quantum dots, color centers in crystals or rare-earth atoms, is of great interest for applications in quantum information processing[3], quantum chemistry[4], and biology[5]. These objects emit light with unique properties including antibunching, absence of ensemble broadening, extreme sensitivity to the local environment, and, in some cases, photon indistinguishability[6–8]. Intrinsically, these sources are usually very faint, making it challenging to harness their emission.

Typical detected photon rates from bare quantum emitters are in the range of $10^4$ to $10^5$ counts per second (cps)[9], only rarely exceeding $10^6$ cps[10–12]. The most fundamental limiting factors for the detected photon rate from quantum emitters are radiative lifetime and photon collection efficiency. By solely increasing the collection efficiency[13–17], detected photon rates from single emitters can exceed $10^6$ cps[13]. Another approach for brightness enhancement hinges on drastically increasing the local density of states (LDOS) in the vicinity of the emitter, leading to a faster photon emission[18]. Dielectric[19–21] and plasmonic[22] resonators are typically used for LDOS enhancement, with alternative approaches relying on slow light[23,24] and metamaterial dispersion[25,26]. Therefore, identifying the structures which combine LDOS engineering with improved collection efficiency[27,28] appears to be the most promising approach for a significant improvement of the single photon emitter brightness.

The LDOS is proportional to the ratio of the resonance quality factor and the volume where the field is localized. In accord with this, LDOS enhancement is typically achieved by either using highly resonant but relatively large (diffraction-limited) dielectric structures or plasmonic nanostructures with lower resonance quality-factors but smaller, sub-diffraction volumes. Dielectric resonators and slow light waveguides enhance the LDOS by increasing the interaction time between the dipole and the emitted field. However, as the quality factor of the dielectric structure increases, the photon storage time becomes an impediment to faster emission and caps the observable lifetime shortening[29]. Alternatively, plasmonic nanostructures ensure broadband (i.e., low-Q) LDOS enhancement and their theoretical potential for lifetime shortening is about two orders of magnitude higher[30]. The shortcoming of plasmonic nanostructures is the high loss resulting from non-radiative quenching of excitation and plasmon absorption in metals.

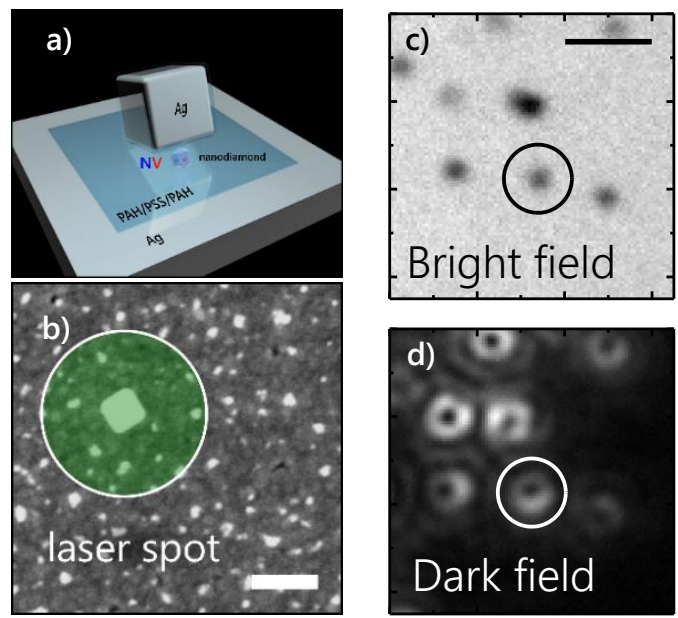

Figure 1. (a) Schematic view of an all-silver nano-patch antenna (NPA): a silver nanocube on silver substrate, with a nanodiamond placed in-between. (b) SEM micrograph of the Ag substrate with randomly dispersed nanodiamonds and a 100 nm Ag nanocube. The scale bar in the pictures is 200 nm. The green circle shows the scale of the focused laser spot used for optical excitation. (c) and (d) Bright field and dark field camera images of the NPA-enhanced emitters. The images of the emitter described in the text are circled.

Nanostructures with gap-surface-plasmon or metal-insulator-metal (MIM) modes strongly alleviate the quenching problem. Despite their proximity to the metal, the coupling rate of the emitter to the gap plasmonic mode is comparable to the quenching rate even as the gap shrinks below 10 nm[31]. Additionally, forming a gap-plasmon resonator by replacing the top metal layer with a nanoparticle leads to a nano-patch antenna (NPA) geometry. This geometry leads to extreme concentration of electromagnetic field, allowing the observation of emitter-plasmon strong coupling at room temperature[32]. The NPAs radiate plasmonic energy on a time scale comparable to the plasmon lifetime[33], which mitigates plasmon absorption. Because of this, NPAs are emerging as the systems of choice for enhancing the brightness of nanoscale emitters[34,35]. Impressive results of single-photon emission in NPAs have been recently demonstrated using colloidal quantum dots (CQDs) with a strong lifetime shortening, antibunching photon statistics and photon counts exceeding 1 Mcps[1]. Nevertheless, due to photobleaching of quantum dots used in that experiment, it was not possible to reach the important saturation regime. The full potential of quantum emitter brightness enhancement by NPAs still remains to be realized. In this work, we use an all-silver NPA to enhance the brightness of a photostable emitter - a nitrogen-vacancy (NV) center[36] in a nanodiamond (see Figure 1(a)) - to almost 5 Mcps in saturation, which to our knowledge is among the highest levels reported so far. This brightness level was achieved by shortening the excited-state lifetime below 1 ns, while avoiding excessive energy loss to quenching. Even higher NV brightness was achieved for NPAs made with epitaxial silver. By observing fluorescence saturation and radiation patterns, we were able to quantify the factors leading to the brightness enhancement.

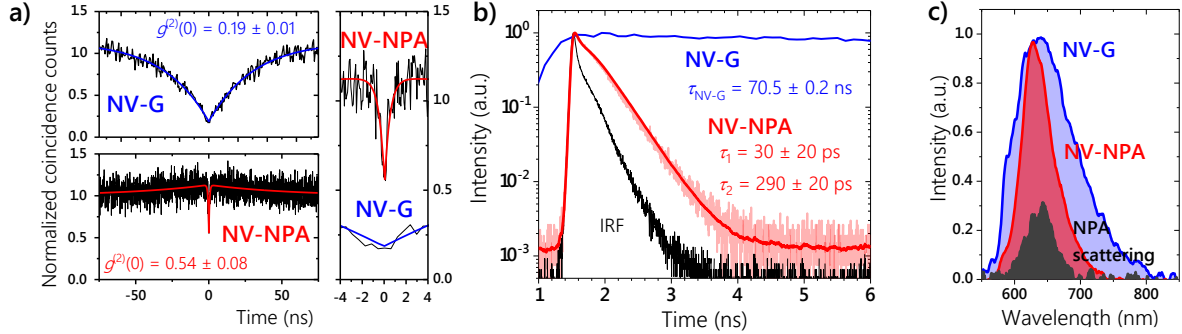

Figure 2. Photophysical characterization of a control nitrogen-vacancy (NV) on a coverslip glass substrate (blue) and NV in an NPA (red). Representative examples of (a) autocorrelation curves, (b) fluorescence decay curves, and (c) spectra of emission and dark field scattering spectrum of the NPA (gray). The IRF curve in b) describes the instrument response function. The fluorescence decay curve of the NV-NPA (light pink) is fitted (red) using a sum of two exponential decays with time constants $\tau_1$ and $\tau_2$ convoluted with the instrument response function.

In our experiment, the NVs were hosted by nanodiamonds with a diameter $d_{ND} = 20 \pm 5$ nm, where each nanodiamond nominally contained one to two NVs. Two samples with NV emitters were used in the study, one coupled to an NPA and another with a glass coverslip substrate. The sample with NPAs was fabricated according to the procedure well documented in prior works[1,37,38] by depositing a 50 nm polycrystalline Ag film and forming an approximately 6 nm thick dielectric spacer through layer-by-layer alternate deposition of poly-allyalamine hydrochloride (PAH) and polystyrene sulfonate (PSS). Nanodiamonds and subsequently 100 nm size single-crystalline nanocubes were randomly dispersed on the spacer layer. This step resulted in the random formation of NPAs with NV centers inside (Figure 1(b)). The NPAs were first imaged using bright field microscopy (Figure 1(c)). Dark field images (Figure 1(d)) reveal scattering profiles from the nanoantennas, which is characteristic for NPA structures with a spacer layer of thickness less than 10 nm[38]. In addition, a similar sample was made using epitaxial silver that exhibits at least 5 times lower losses than the polycrystalline film at the wavelengths of interest. Due to lower losses, a further increase in brightness could be observed as discussed below (see also Supplementary Section V).

The control sample was fabricated by randomly dispersing the nanodiamonds on a glass coverslip substrate with refractive index $n = 1.525$. In the control experiment, we have measured the photophysical characteristics of 14 single NV centers. We selected one representative emitter from both the NPA-enhanced set and the control set (see Supplementary Section V) and compared those emitters. From now on, the representative NPA-enhanced and coverslip glass-based sources will be referred to as NV-NPA and NV-G, respectively.

The NV-NPA emitter was analyzed using an air objective with a numerical aperture of 0.9. Antibunching behavior was characterized by the second order autocorrelation function $g^{(2)}_{\text{NV-NPA}}(t)$ measured under continuous excitation. The value of $g^{(2)}_{\text{NV-NPA}}(0) = 0.54 \pm 0.08$ points towards the presence of two NVs inside the NPA (Figure 2(a)). Background emission from the planar silver film was found negligible and not likely to significantly contribute to $g^{(2)}_{\text{NV-NPA}}(0)$. Using pulsed excitation, we characterized the fluorescence lifetime by measuring the fluorescence decay (Figure 2(b)). The NV-NPA fluorescence decay consists of two components with characteristic times $\tau_1 = 30 \pm 20$ ps and $\tau_2 = 290 \pm 20$ ps. These time constants were determined by convolution of the sum of two exponential decays with the instrument response function (IRF) having a full width half maximum of 96 ps. The slower component is attributed to the NV emission, as $\tau_2$ is consistent with the emission rate obtained from simulation (see Supplementary Section III). The antibunching of the faster decay component cannot be measured in our experimental setting. Therefore, it cannot be safely attributed to the NV emission, rather than to the light produced either by the metal nanostructure[39] or the polymer layers.

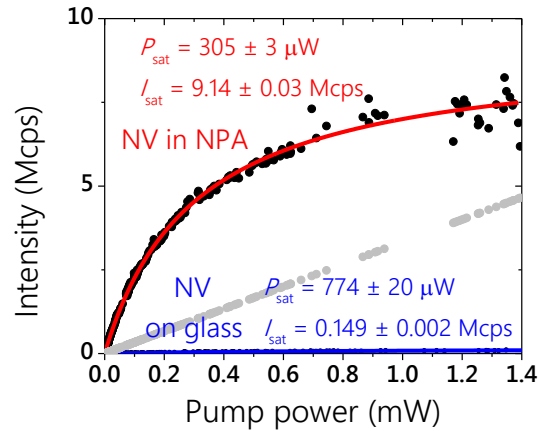

Figure 3. Emission intensity saturation curve for a nanodiamond under an NPA (red) and a reference nanodiamond on glass substrate (blue). Linear component of the saturation (light gray) has been subtracted from the NPA data.

The observed fluorescence lifetime for nanodiamond-based NVs is remarkably shorter compared to what has been observed for NVs in small nanodiamonds or near the diamond-air interface, which typically exhibit fluorescence lifetimes in excess of 10 ns[40,41]. NVs in nanodiamonds have also been reported to suffer from low quantum efficiency[42], which impedes fluorescence lifetime shortening. So far, despite the implementation of both photonic and plasmonic LDOS enhancement techniques, the emission lifetimes obtained with NVs have failed to fall below 1 ns. In our experiment, large LDOS and high outcoupling efficiency of NPAs resulted in an ultrabright antibunched emitter with a sub-ns fluorescence lifetime. By comparing the spectra of nanoantenna scattering and NV-NPA emission (Figure 2(c)), we conclude that the enhanced emission indeed resulted from coupling to the mode of the NPA. The fluorescence saturation curve was fitted using a saturable and a linear component. The linear component is plotted on Figure 3 with gray dots and could possibly be attributed to the same source as the fast component of the fluorescence decay. The saturable component (red curve) yields $I_{\text{NV-NPA}}^{\text{sat}} = 9.14 \pm 0.03$ Mcps, corresponding to $4.57 \pm 0.02$ Mcps per single NV center. Finally, the saturating laser power (measured before coupling to the objective) was $P_{\text{NV-NPA}}^{\text{sat}} = 0.305 \pm 0.030$ mW.

We note that by using an epitaxial silver substrate instead of the polycrystalline silver substrate, we observed emission with anti-bunching and lifetime characteristics comparable to the NV-NPA. For a representative NPA-enhanced emitter on epitaxial silver substrate, we found $g^{(2)}(0) = 0.58 \pm 0.05$ and $\tau_2 = 540 \pm 20$ ps. However, due to lower losses of the epitaxial silver, the saturated emission was about 4 times brighter than that of NV-NPA: $I^{\text{sat}} = 39 \pm 5$ Mcps, corresponding to 19 Mcps $\pm$ 2 Mcps per single NV center, assuming again two emitters in the NPA. These brightness levels demonstrate the potential of NPAs for enhancing the brightness of quantum emitters. The characterization data for additional NPA-enhanced emitters on polycrystalline and epitaxial silver substrates is shown in the Supplementary Section V.

The NV-G source was analyzed using an oil objective with a numerical aperture of 1.49 in the total internal reflection mode. The photon purity was characterized by an autocorrelation at zero delay of $g_{\text{NV-G}}^{(2)}(0) = 0.19 \pm 0.01$, limited by the background fluorescence from the coverslip glass. The measured fluorescence lifetime was 70.5 $\pm$ 0.2 ns, 240 times longer than in the case of the NV-NPA. We observed saturated photon counts of $I_{\text{NV-G}}^{\text{sat}} = 149 \pm 2$ kcps and a saturating laser power of $P_{\text{NV-G}}^{\text{sat}} = 0.774 \pm 0.020$ mW. Table 1 summarizes the photophysical properties of the NV-NPA and NV-G photon sources.

|  | $\gamma^{\text{tot}}$ (ns$^{-1}$) | $I^{\text{sat}}$ (Mcps) | $P^{\text{sat}}$ (μW) | $c^{\text{exc}}$ (GHz/mW) |
|---|---|---|---|---|
| NV-NPA | $(0.29 \pm 0.02)^{-1}$ | $4.57 \pm 0.02$ | $305 \pm 30$ | $11.3 \pm 0.2$ |
| NV-G | $(70.5 \pm 0.2)^{-1}$ | $0.149 \pm 0.002$ | $774 \pm 20$ | 0.018 |
| Ratio | 240 | 30 | 0.4 | 630 |

Table 1. Comparative summary of the total decay rate $\gamma^{\text{tot}}$, emission intensity at saturation $I^{\text{sat}}$, saturating laser power $P^{\text{sat}}$ and the specific excitation rate $c^{\text{exc}} = \gamma^{\text{tot}}/P^{\text{sat}}$ for the NPA emitter (NV-NPA) and the reference emitter on coverslip substrate (NV-G).

The brightness increase due to the NPA can be quantified in two ways. An increase of 30 times in the saturated photon counts was observed for a single NV-NPA emitter compared to the NV-G emitter. Additionally, we registered a brightness increase of 70 times of the NV-NPA compared to an average single NV on a glass substrate measured with the same objective configuration, i.e. an air objective with numerical aperture of 0.9 (see Supplementary Section IV for full statistics).

Despite a drastic lifetime shortening compared to the NV-G, the NV-NPA emission is saturated at a comparable excitation laser power. Therefore, the excitation rate at constant pump power is also strongly enhanced by the NPA. Specifically, the concentration of the pump field in the NPA results in NV excitations of the NV-NPA that are 630 times faster compared to the NV-G, for the same incident pump power (see Table 1). Enhancement of both emission and excitation rates is a consequence of multiple broad resonances present in NPAs[43]. It has been previously shown that at high incident pump powers, the properties of the optical modes in NPAs were altered due to structural deformation[44]. Strong local enhancement of pump intensity in NPAs allows to obtain the same excitation rate at a reduced total incident pump power. Therefore, the local pump enhancement is important to guarantee a high emitter brightness, without degrading the antenna properties. In the future, these resonances could be engineered[45] to precisely match the absorption and emission frequencies of narrow-band emitters, yielding an even larger intensity enhancement and better photon purity[46].

The saturated detected count rate and the calculated total efficiency of the NV-NPA depend on the efficiency of the microscope setup. To obtain an instrument-independent quantity, we calculated the NPA's photon efficiency, i.e. the number of photons emitted into the far field per NV excitation event. The photon efficiency can either be calculated from the assumed microscopic properties of the photon source or deduced from the observed total fluorescence intensity. We use both approaches to estimate, respectively, the upper ($\eta^{\text{ph, sim}}$) and lower ($\eta^{\text{ph, meas}}$) bounds for the photon efficiency of NV-NPA and NV-G photon sources. These quantities are summarized in Table 2.

The upper bound for photon efficiency is obtained from full-field simulations. We define the simulated photon efficiency as

$$\eta^{\text{ph, sim}} = \frac{\gamma^{\text{ff}}}{\gamma^{\text{ff}} + \gamma^{\text{loss}} + \gamma^{\text{nr}}} \ , \tag{1.1}$$

where $\gamma^{\text{ff}}$ is the rate of photon emission into the far-field by an NV center with unity quantum yield, $\gamma^{\text{loss}}$ is the total rate of dissipation to heat consisting of quenching and plasmon absorption and $\gamma^{\text{nr}}$ is the intrinsic rate of non-radiative decay. For the NV-G case, there are no losses in the metal, so the photon efficiency is identically equal to the quantum yield and limited by $\gamma^{\text{nr}}$. For the NV-NPA case, the photon efficiency is, in contrast, dominated by losses to the metal, while the nonradiative rate is bound by the total measured decay rate of NVs on glass and is therefore negligible.

In order to calculate $\gamma^{\text{ff}}$ and $\gamma^{\text{loss}}$, we first obtain the electromagnetic field distributions near the emitters. Figures 4(a) and (b) show the plots of the normalized electric field intensities, with a several orders of magnitude enhancement in the near-field of the NPA. The observed lifetime shortening of 240 times is in good agreement with a 200-fold shortening obtained in the simulation. By calculating the fraction of the total emitted power that is lost in the metal layers, we estimate the photon efficiency of the NV-NPA photon source to be $\eta^{\text{ph, sim}}_{\text{NV-NPA}} \approx 77\%$ (see Supplementary Section III). The lower bound on $\gamma^{\text{nr}}$ is obtained from the measurement of the fluorescence lifetime (see Supplementary Section VI). This estimate leads to $\eta^{\text{ph, sim}}_{\text{NV-G}} \approx 45\%$.

The lower bound on the photon efficiency $\eta^{\text{ph, meas}}$ is calculated based on the measured total efficiency $\eta^{\text{tot}}$ using the following simple relation:

$$\eta^{\text{tot}} = \eta^{\text{ph, meas}} \eta^{\text{col, meas}} \eta^{\text{setup}}, \tag{2}$$

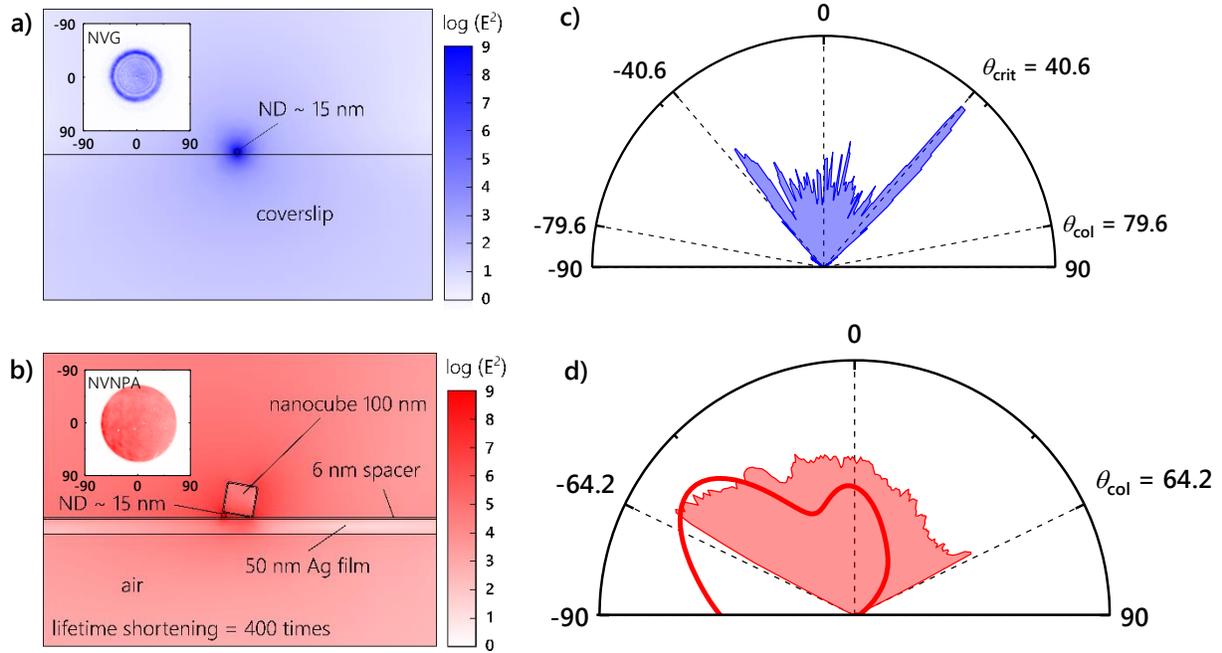

Figure 4. Logarithmic electric field intensity for the NV-G (a) and the NV-NPA (b). Insets show the corresponding intensity distribution measured in the back focal plane (BFP). Radiation patterns determined from BFP measurements for the NV-G (c) and the NV-NPA (d). Red solid line in d) represents the result of a numerical simulation (see Supplementary Section III)

where $\eta^{\text{col, meas}}$ is the collection efficiency of the emitted photons by an objective inferred from measurement and $\eta^{\text{setup}}$ is the detection efficiency of the scanning confocal microscope, which includes the losses incurred at all the optical components and detector efficiency. $\eta^{\text{setup}} \approx 14\%$ is directly measured by using a collimated laser beam of known power at 638 nm wavelength flooding the back aperture of the objective and reflected by a mirror placed at the sample location (for more details, see Supplementary Section II). We assume that the setup efficiency for both the NV-G and NV-NPA sources was the same, i.e. that the oil objective (NA = 1.45) and air objective (NA = 0.9) have the same transmission coefficient in the red region of the visible range. $\eta^{\text{col, meas}}$ is determined based on the results of the back-focal plane measurements (see insets in Figure 4 (a) and (b)). The NV-G source is expected to radiate mostly into the glass with an emission angular pattern concentrated around the critical angle. A simple analytical estimate predicts the fraction of the total emitted power into the glass to be 82% for the in-plane dipole and 85% for the out-of-plane dipole. The results of this model agree reasonably well with the emission's measured angular distribution (see Figure 4 (c)). The observed deviation from the model is most likely due to objective-induced aberrations. Therefore, we assume that the collection efficiency for the NV-G source is 85%. On the other hand, the emission pattern of the NV-NPA source should vary widely depending on the exact relative position of the nanodiamond and the nanocube. In the experiment, we observe an essentially non-directional pattern within the NA = 0.9 solid angle. Assuming that the emission is roughly isotropic in the upper half space, we estimate $\eta^{\text{col, meas}}_{\text{NV-NPA}} \approx 50\%$. With these numbers, we estimate the lower bound photon efficiencies at $\eta^{\text{ph, meas}}_{\text{NV-G}} = 9\%$ and $\eta^{\text{ph, meas}}_{\text{NV-NPA}} = 1.8\%$. Correspondingly, the NV-NPA is emitting at least 60 million photons per second into the far-field at saturation.

$\eta_{\text{NV-NPA}}^{\text{ph, low}}$ is probably dominated by plasmon losses and quenching, both of which can be improved by employing better plasmonic materials. For example, epitaxial silver is superior to the conventional amorphous silver film both in terms of surface roughness, bulk ohmic losses[47,48] and speed of degradation. We observe a photon efficiency that is 5 times lower for the NV-NPA compared to the NV-G. We attribute this drop to the imperfections of the silver film, such as nanoscale roughness and chemical degradation of the topmost layer, which enhance optical losses. For the representative NPA-enhanced emitter on epitaxial silver, using the above method, we estimate $\eta^{\text{ph, meas}} \approx 15\%$ and that about 250 million photons per second are emitted into the far-field at saturation.

|  | $\eta^{\text{tot}}$ (%) | $\eta^{\text{ph, meas}}$ (%) | $\eta^{\text{ph, sim}}(z)$ (%) | $\eta^{\text{ph, sim}}(x)$ (%) | $\eta^{\text{ph, sim}}(y)$ (%) |
|---|---|---|---|---|---|
| NV-NPA | 0.13 ± 0.01 | 1.8 | 77 | 12 | 85 |
| NV-G | 1.05 ± 0.01 | 9 | 45 | | |

Table 2. Summary of NV-G and NV-NPA efficiency characterization.

In summary, we demonstrate an ultrabright single-photon source utilizing a nitrogen-vacancy center in a nanodiamond in the gap of an all-silver nano-patch antenna. The all-silver structure and the emitter are stable at excitation power densities on the order of $10^4$ W/cm$^2$. In the gap of the nano-patch antenna, the fluorescence lifetime is typically shortened by more than 200 times compared to the average value for similar emitters on a coverslip substrate. For representative NPAs on polycrystalline and epitaxial silver films, the saturated single-photon detected rate reaches 4.57 and 19 million counts per second, respectively, whereas the photon production rates are $6 \cdot 10^7$ and $2.5 \cdot 10^8$ photons per second, respectively. The NPAs show a greatly improved excitation efficiency indicating that they are well suited for enhancing weakly absorbing emitters such as diamond color centers. We note that these results could be further improved by employing passivated epitaxial silver substrates, deterministic positioning of the nanocube with respect to the emitter[49] and control over emitter dipole orientation. These results show strong promise of silver nano-patch antennas for engineering ultrabright plasmon-assisted emission from single fluorescent objects. Coupling different emitter types to similar nanogap structures may be of great interest for single-molecule spectroscopy[50], multi-photon microscopy[51], nanomagnetometry[52], and photonic quantum technologies[53]. In particular, it could foster new emerging fields of ultrafast single-molecule nanoscopy[54], probing weak and forbidden electronic transitions[55,56], three-dimensional single-molecule nuclear magnetic resonance[57] and cluster state engineering[58].

**Methods.**

**Sample fabrication.** The coverslip sample was prepared by coating the coverslip surface with one PAH monolayer and dropcasting a diluted water solution of 20 nm nanodiamonds for 10 seconds (Adamas Nano) before rinsing the surface with water. Polycrystalline 50 nm silver films were deposited on Si substrate with a 10 nm Ti adhesion layer. Epitaxial silver films were 35 nm thick and deposited directly on Si substrate. Three self-assembled monolayers (PAH/PSS/PAH) were deposited by alternately dipping the samples with the silver films into the corresponding solutions for 5 min. A diluted water solution of 20 nm nanodiamonds was then dropcast onto the PAH surface. Finally, a 20X diluted water/ethanol solution of 100 nm crystalline nanocubes (Nanocomposix) was dropcast on top of the nanodiamonds.

**Characterization.** All the optical characterization was performed using a home built scanning confocal microscope with a 50 μm pinhole based on a commercial inverted microscope body (Nikon Ti-U). The optical pumping in all the experiments except the lifetime measurement was administered by a DPSS 532 nm laser. Lifetime characterization was performed using a 514 nm fiber-coupled diode laser with a nominal 100 ps pulse width and adjustable repetition rate in the 2 – 80 MHz range (BDL-514-SMNi, Becker&Hickl). The excitation beam was reflected off a 550nm longpass dichroic mirror (DMLP550L, Thorlabs) and a 550 nm long-pass filter (FEL0550, Thorlabs). Two avalanche detectors with a 30 ps time resolution and 35% quantum efficiency at 650 nm (PDM, Micro-Photon Devices) were used for single-photon detection during scanning, lifetime and autocorrelation measurements. An avalanche detector with 69% quantum efficiency at 650 nm (SPCM-AQRH, Excelitas) was used for saturation measurements. Time-correlated photon counting was assured by an acquisition card with a 4 ps internal jitter (SPC-150, Becker & Hickl). The measured value of $g^{(2)}_{\text{NV-NPA}}(0)$ was influenced by the time resolution of the instrument, which in the autocorrelation experiment was $\delta\tau_{\text{instr}} = 80 \text{ ps}$. The resulting increase in $g^{(2)}_{\text{NV-NPA}}(0)$ can be estimated as $\delta g^{(2)}_{\text{instr}} \approx 1 - (1-e^{-\zeta})/\zeta = 0.06$, where $\zeta = \delta\tau_{\text{instr}}/2\tau_2 \approx 0.12$ is the normalized time resolution of the autocorrelation experiment.

**Simulation.** All full-field simulations were performed using Comsol Multiphysics. The simulation domain was a cube with a side size of 1.8 μm for NV-NPA and 2.4 μm for NV-G surrounded by a 400-nm-thick perfect matching layer (PML). The optical emitter was modeled as an AC current density inside a 2-nm-diameter sphere enclosed by a 15-nm-diameter diamond shell. The emitter's wavelength was fixed at 675 nm. The emitter had a vertical dipole orientation and was placed between a 50 nm Ag layer and an Ag nanocube. The nanocube was modeled having 8 nm curvature radius at its corners and covered in 3 nm PVP layer with $n = 1.4$. One ridge of the nanocube's bottom facet was in full contact with the spacer, while the middle of the opposite ridge of the bottom facet was supported by the nanodiamond. The spacer layer was modeled as an isotropic dielectric with $n = 1.5$.

Simulated values in Table 1 are computed assuming an out-of-plane dipole orientation for the NV-NPA, an in-plane orientation for the NV-G emitter and an intrinsic quantum yield of 1 for both emitters. Vertical dipole orientation for the NV-NPA is assumed dominant because of the relatively weak Purcell effect expected for the two in-plane orientations. An in-plane orientation was chosen for the NV-G emitter because its saturated emission intensity is somewhat below average (see Supplementary Section III).

The authors acknowledge O. Makarova for the AFM measurement of polycrystalline silver RMS roughness as well as I. Aharonovich, M. Mikkelsen, A. Akimov, V. Vorobyev and S. Bolshedvorsky for useful discussions. This work was partially supported by ONR-DURIP Grant No. N00014-16-1-2767, and DOE Grant DE-SC0017717.

# Supplementary information

# Ultrabright room-temperature single-photon emission from nanodiamond nitrogen-vacancy centers with sub-nanosecond excited-state lifetime

S. Bogdanov[1,2], M. Shalaginov[1,2], A. Lagutchev[1,2], C.-C. Chiang[1,2], D. Shah[1,2], A.S. Baburin[3,4], I.A. Ryzhikov[3,5], I.A. Rodionov[3,4], A. Boltasseva[1,2] and V.M. Shalaev[1,2]

[1]*School of Electrical and Computer Engineering, Purdue University, West Lafayette, IN 47907, USA*
[2]*Birck Nanotechnology Center and Purdue Quantum Center, Purdue University, West Lafayette, IN 47907, USA*
[3]*FMNS REC, Bauman Moscow State Technical University, Moscow, 105005, Russia*
[4]*Dukhov Research Institute of Automatics, Moscow, 127055, Russia*
[5]*Institute for Theoretical and Applied Electromagnetics RAS, Moscow, 125412, Russia*

## I. Substrate fabrication

### a. Ag substrate deposition

The polycrystalline silver substrate was deposited using an e-beam evaporator (Leybold) at a base pressure of $2 \cdot 10^{-6}$ Torr. First, an adhesion layer of Ti (10 nm) was deposited on <100> Si substrate, then a 50 nm layer of Ag. The sample spent 10-15 min in air before the deposition of the spacer layer. A control sample underwent the same deposition procedure and was then characterized by ellipsometry to retrieve the optical properties of the silver substrate. The dielectric permittivity data is plotted on Supplementary Figure 1. The wavelength used in simulations is 675 nm and the dielectric permittivity of the substrate at this wavelength is $\varepsilon = -20 + i0.6$.

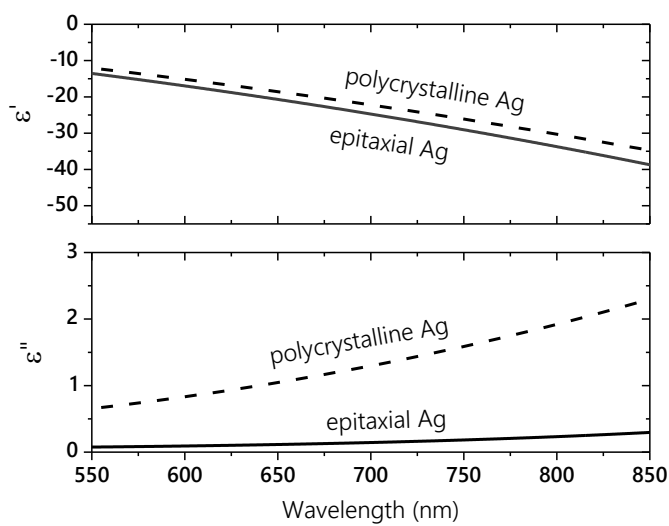

Supplementary Figure 1. Dielectric permittivity constants of polycrystalline and epitaxial Ag obtained from ellipsometry measurements.

The epitaxial silver substrate was deposited using an e-beam evaporator (Angstrom Engineering) at a base pressure of $3 \cdot 10^{-8}$ Torr. 35 nm layer of Ag was deposited on <111> Si substrate without any adhesion layer. The epitaxial silver films were deposited at the BMSTU Nanofabrication Facility (Functional Micro/Nanosystems, FMNS REC, ID 74300) and spent 2 weeks in air before NPA fabrication and measurements. A control sample underwent the same deposition procedure and was then characterized by ellipsometry to retrieve the optical properties of the epitaxial silver substrate. The dielectric permittivity of the epitaxial silver substrate at 675 nm wavelength is $\varepsilon = -22.5 + i0.12$.

### b. Substrate roughness

RMS roughness of the epitaxial silver film (as fabricated) was 0.18 nm measured over $2.5 \times 2.5$ μm², substantially less than the 2.6 nm RMS roughness of poly-silver substrates (see Supplementary Figure 2). Even in the case of epitaxial Ag substrate, surface roughness could lead to the formation of ultrasmall cavities[S1] in the region of contact between the cube and the substrate. The formation of such ultrasmall cavities could explain the existence of the faster fluorescence decay component[S2].

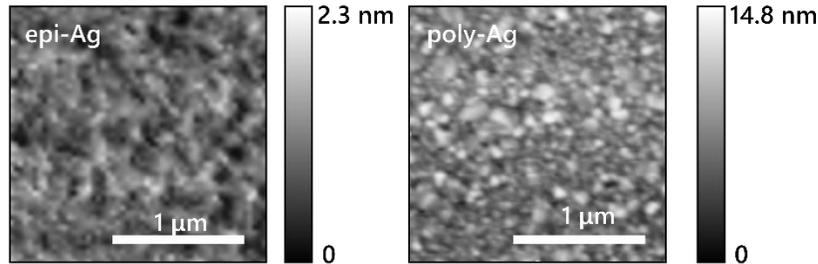

Supplementary Figure 2. AFM scans of epi- and poly-Ag film surfaces. RMS roughness values are 0.18 nm and 2.6 nm respectively measured over $2.5 \times 2.5$ μm².

### c. Spacer deposition

The spacer layer was formed by deposition of individual self-assembled monolayers following a standard procedure described in detail elsewhere[S3,S4] and shortly summarized as follows. First, the silver substrate was dipped into a 0.003 monomol/L poly-allyalamine (PAH) solution with 1 M of NaCl for 5 min, rinsed with water, then immersed into a 1M NaCl solution for 30 seconds. The procedure was then repeated for the deposition of a polystyrene sulfonate layer (PSS) and the final PAH layer. The compounded thickness of three monolayers was $6 \pm 2$ nm as determined by ellipsometry.

## II. Setup efficiency

Setup efficiency is defined as the fraction of photons emitted within the objective's collection angle that are registered by the detector. A collimated 638 nm laser beam was coupled into the BD objective, and we measured its power at the sample position using a sensitive power meter (PM16-130, Thorlabs). Then, a mirror with 95% nominal reflectivity was installed at the sample position and the reflected laser light was registered by the single-photon detector that was used in the saturation curve measurements (SPCM-AQRH, Excelitas). We then divided the measured photon count rate at the single-photon detector by the photon flux measured by the power meter at the sample location. The measurement yielded a setup efficiency of $14 \pm 1\%$.

We check this measurement by numerically compounding nominal optical losses at all the elements in the fluorescence path. The losses include the following components: objective transmission, reflections at the mirrors, dichroic mirror transmission, transmission of confocal telescope, consisting of two lenses and one pinhole, transmission of detector's lens and detector quantum efficiency. Accounting for these losses, we obtain an "ideal" setup efficiency of 20 ± 1%. The measured value is in good agreement with our estimate. The discrepancy may be due to the imperfections in the microscope alignment, optical aberrations and accidental dust particles on the optical components.

## III. Simulations of emission properties

The numerical electro-magnetic calculations were performed in COMSOL 5.3, Wave Optics Module. Optical permittivity data of polycrystalline silver from Supplementary Figure 1 was used to simulate the emission of the NV-NPA. The simulation domain represents a cube with a side of 1.8 µm or 2.4 µm for NVNPA or NVG configurations, respectively. In both cases the domains were truncated with a standard 400-nm-thick PML layer. The dipole emitter at 675-nm-wavelength was introduced as a current oscillating inside a 1-nm-radius sphere. The total decay rate $\gamma$ is calculated as the surface integral of total power flow **P** through a 3-nm-radius spherical surface $\Omega_{in}$ encapsulating the emitting dipole and situated entirely within the nanodiamond volume (see Supplementary Figure 3): $\gamma \propto \int_{\Omega_{in}} \mathbf{P} \cdot \mathbf{dS}$. The loss rate $\gamma^{loss}$ is calculated as the total work performed by the electric field on the free charges in the metal regions occupying the volume $V_m$ : $\gamma^{loss} = \int_{V_m} \mathbf{j} \cdot \mathbf{E} dV$.

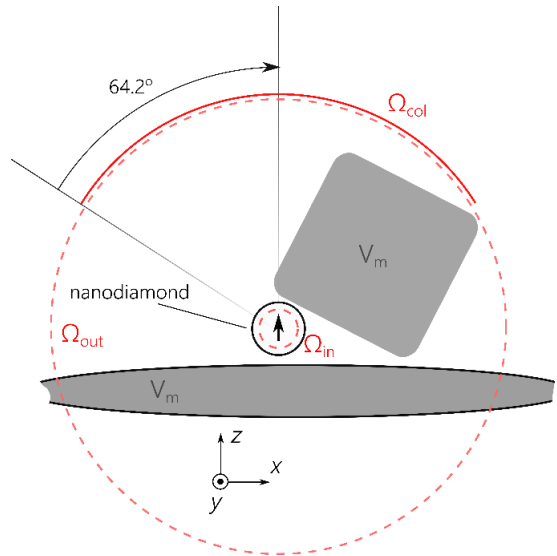

Supplementary Figure 3. Schematic illustration of integration volumes and surfaces used for simulating efficiency parameters of the NVNPA.

All the numerically obtained power flows are normalized by the power dissipated by an NV center in bulk diamond corresponding to a known bulk decay rate of $\gamma^{bulk} = (12.8 \text{ ns})^{-1}$. The simulated rates can therefore be expressed in units of ns$^{-1}$. Finally, the collection efficiency $\eta^{col}$ is calculated as the ratio of the far-field electric fields squared integrated over a spherical outer surface $\Omega_{out}$ (0.7-µm-radius sphere for NV-NPA and 1-µm for NVG) and the portion of that surface $\Omega_{col}$ corresponding to the collection solid angle of the relevant objective: altitude angles of 64° (0.9 NA) and 79.6° (1.49 NA) for NV-NPA and NV-G, respectively). The far-field electric fields were calculated using a standard near-field to far-field transformation.

The summary of numerically obtained quantities $\gamma$, $\gamma^{\text{loss}}$, $\eta^{\text{col}}$ for basis dipole orientations in NV-NPA and NV-G cases are presented in Supplementary Table 1. The quantity $\gamma^{\text{ff}}$ used in the main text represents the difference $\gamma - \gamma^{\text{loss}}$.

| Emitter and orientation | $\gamma$ (ns$^{-1}$) | $\gamma^{\text{loss}}$ (ns$^{-1}$) | $\eta^{\text{col}}$ (%) |
|---|---|---|---|
| NV-NPA ($z$-dipole) | 0.44$^{-1}$ | 1.9$^{-1}$ | 56 |
| NV-NPA ($x$-dipole) | 5.5$^{-1}$ | 22.3$^{-1}$ | 56 |
| NV-NPA ($y$-dipole) | 72.2$^{-1}$ | 76.4$^{-1}$ | 56 |
| NV-G ($z$-dipole) | 84.5$^{-1}$ | 0 | 85 |
| NV-G ($x$-dipole) | 147$^{-1}$ | 0 | 82 |

Supplementary Table 1. Summary of simulated rates of dipole emission, photon loss to metal as well as collection efficiency.

We performed an estimate of the collection efficiency for the NV-G source using semi-analytical calculations. For a single dipole emitter with in-plane (x,y) or perpendicular (z) orientation placed at a distance $h$ above oil/glass planar interface, the corresponding $\eta^{\text{col}}$ can be calculated using dyadic Green function formalism[S5].

$$\eta^{\text{col}} = \int_{\varepsilon_{\text{sup}}^{1/2}}^{\infty} p_\kappa d\kappa \Bigg/ 1 + \int_0^{\infty} p_\kappa d\kappa, \qquad \text{S(1)}$$

where depending on dipole orientation, wave-vector density is expressed as

$$p_\kappa^{(z)} = \frac{3}{2} \frac{1}{\varepsilon_{\text{sup}}^{3/2}} \text{Re}\left\{ \frac{\kappa^3}{\kappa_{z,\text{sup}}(\kappa)} r^{\text{P}}(\kappa) e^{2ik_0 \kappa_{z,\text{sup}}(\kappa) h} \right\}, \qquad \text{S(2)}$$

$$p_\kappa^{(x,y)} = \frac{3}{4} \frac{1}{\varepsilon_{\text{sup}}^{1/2}} \text{Re}\left\{ \frac{\kappa}{\kappa_{z,\text{sup}}(\kappa)} \left[ r^{\text{s}}(\kappa) - \frac{\kappa_{z,\text{sup}}^2(\kappa)}{\varepsilon_{\text{sup}}} r^P(\kappa) \right] e^{2ik_0 \kappa_{z,\text{sup}}(\kappa) h} \right\}. \qquad \text{S(3)}$$

In equations S(1) - S(3), $\kappa = k_{x,y}/k_0$; $\kappa_{z,\text{sup}}(\kappa) = k_{z,\text{sup}}/k_0 = \left(\varepsilon_{\text{sup}} - \kappa^2\right)^{1/2}$; $\varepsilon_{\text{sup}}$, $\varepsilon_{\text{sub}}$ are superstrate and substrate relative permittivities, i. e. $\varepsilon_{\text{sup}} = 2.3$ and $\varepsilon_{\text{sub}} = 2.16$; $r^{\text{P}}$ and $r^s$ are conventional Fresnel coefficients: $r^{\text{P}} = \dfrac{\varepsilon_{\text{sub}} \kappa_{z,\text{sup}} - \varepsilon_{\text{sup}} \kappa_{z,\text{sub}}}{\varepsilon_{\text{sub}} \kappa_{z,\text{sup}} + \varepsilon_{\text{sup}} \kappa_{z,\text{sub}}}$, $r^s = \dfrac{\kappa_{z,\text{sup}} - \kappa_{z,\text{sub}}}{\kappa_{z,\text{sup}} + \kappa_{z,\text{sub}}}$. The integrals were numerically evaluated by using an adaptive Gauss–Kronrod quadrature method. Semi-analytically calculated values of $\eta^{col}$ are 82% and 85% for (x,y) and z dipole orientations, correspondingly, which agrees well with the values obtained from numerical simulations.

## IV. Statistics of reference emitters: single NVs on coverslip glass

In the control experiment, we have measured the photophysical properties of 15 nanodiamonds containing single NVs which were dispersed on a coverslip glass substrate. Only emitters with antibunching characterized by $g^{(2)}(0) < 0.5$ were selected to be part of the statistics. The summary of the measurement results is shown in Supplementary Figure 4. The spread in fluorescence lifetime arise from two separate causes. First, the radiative decay rates depend on the dipole orientation (see Supplementary Section III). Second, the variations in quantum yield, which are known to be wide in commercial nanodiamonds[S6] (see Supplementary Section VI for more details on estimating quantum yield), also contribute to the observed lifetime. The spread in saturation intensity is due to variations in dipole orientation, which affects the radiative decay rate. The saturating pump intensities are the most broadly distributed because they are affected by both the fluorescence lifetime and random dipole orientations with respect to the pump polarization direction.

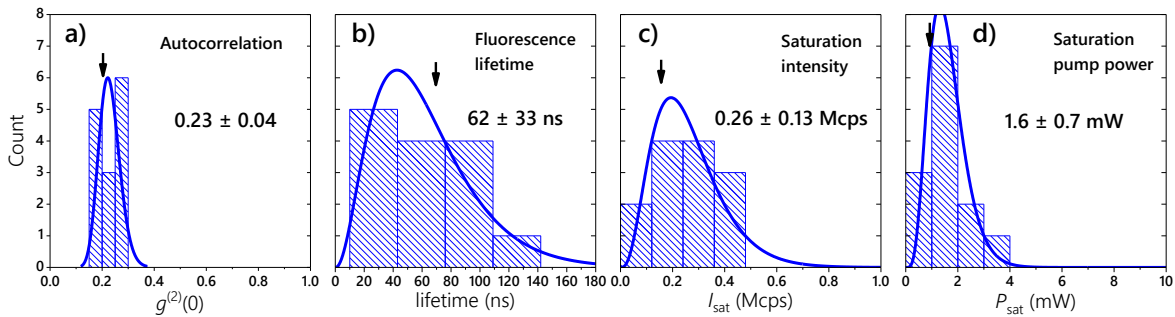

Supplementary Figure 4. Statistical distributions of the photophysical properties of NV centers in nanodiamonds dispersed on glass coverslip and characterized with an oil objective with NA = 1.49 in the TIRF configuration. (a) Antibunching at zero delay in the autocorrelation function. (b) Fluorescence lifetime. (c) Saturated emission intensity. (d) Saturating pump power. Solid lines are gamma distribution fits to the measured data.

Black arrows indicate the values corresponding to the representative emitter chosen as the reference NVG source described in the main text. For all the measured quantities, these values are close to the maxima of the probability distributions, which make this emitter representative of the whole ensemble.

# V. Additional NPA-enhanced emitters

## a. on polycrystalline silver substrate

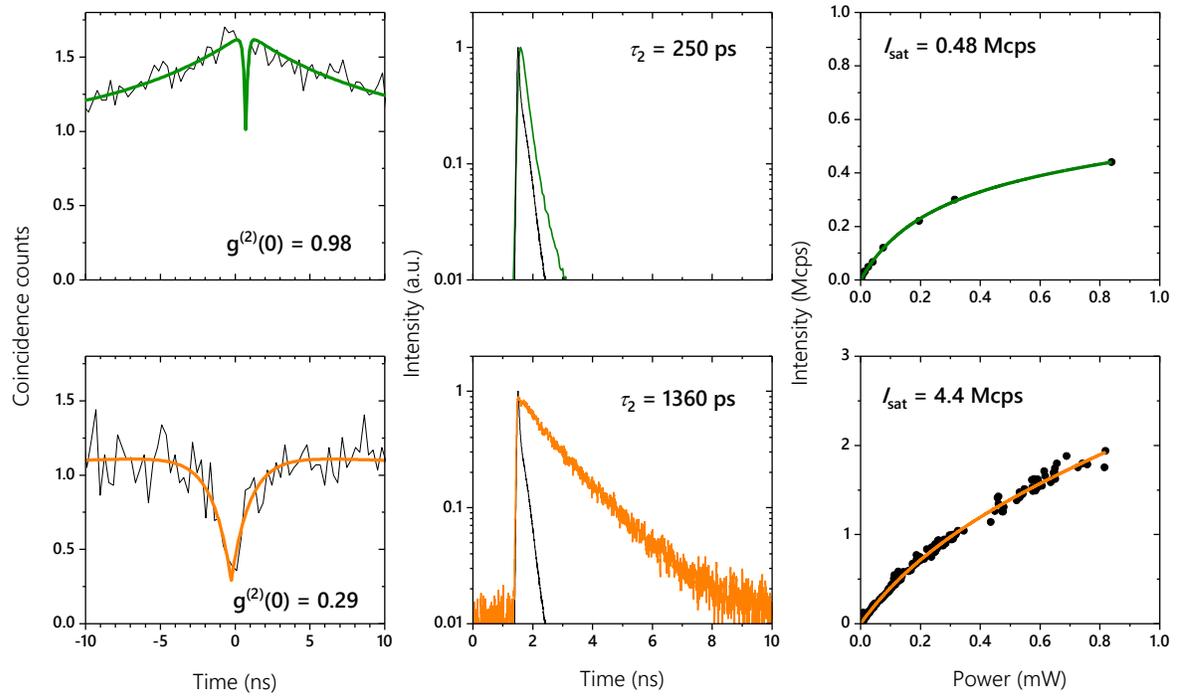

Supplementary Figure 5. Autocorrelation curves (left column), fluorescence decays (middle column) and saturation curves (right column) for two additional NV center emitters enhanced by NPA structures, similar to the NV-NPA emitter described in the main text. Both emitters were assembled on polycrystalline silver substrate.

### b. on epitaxial silver substrate

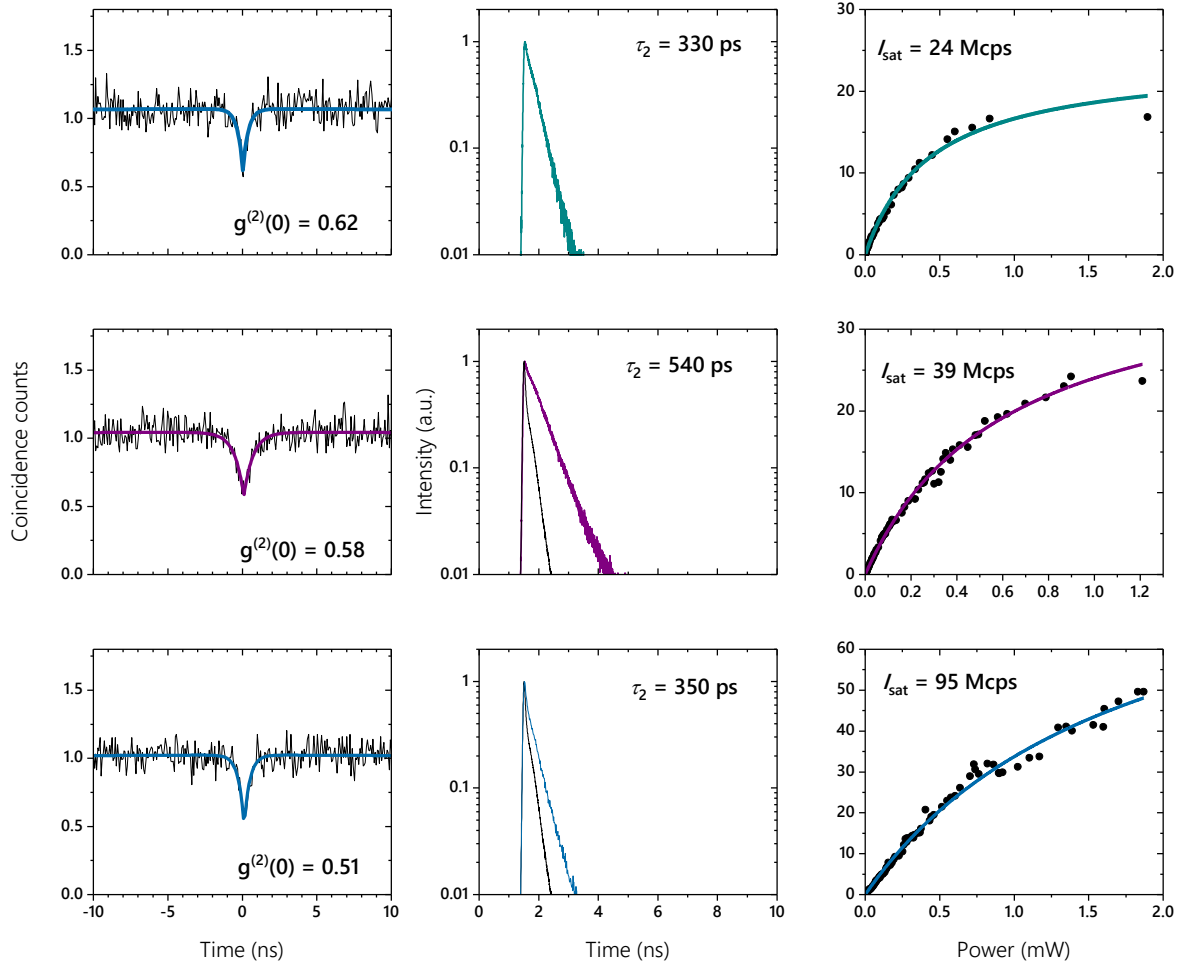

Supplementary Figure 6. Autocorrelation curves (left column), fluorescence decays (middle column) and saturation curves (right column) for three NV center emitters enhanced by NPA structures assembled on epitaxial silver substrate. A substantial increase in saturated intensity can be noticed compared to the emitters assembled on polycrystalline silver substrate.

## VI. NV quantum yield

We estimate the NV quantum yield from measurements of fluorescence lifetime. This method relies on several assumptions. First, we assume NV's intrinsic lifetime in bulk diamond to be constant $\tau_{\text{bulk, i}} = 12.8$ ns. Second, we assume that the relation between the lifetime of an emitter in bulk diamond and emitter in a nanodiamond obeys the following analytical expression[S7]:

$$\frac{\tau_{\text{ND, 0}}}{\tau_{\text{bulk, 0}}} = n \left( \frac{n^2 + 2}{3} \right)^2 \approx 17 \qquad \text{S(4)}$$

For the emitters on glass, an additional lifetime shortening is given by our numerical simulations: $\frac{\tau_{\text{g, 0}}}{\tau_{\text{ND, 0}}} \approx 0.7$. Here, the index 0 refers to the values which do not account for the non-radiative decay processes. As all the nanodiamonds in our study are highly subwavelength, the variations in their sizes and shapes are not expected to influence these lifetime values. Therefore, we can deduce the quantum yield of our NVs by comparing the measured lifetime on glass to the theoretical lifetime $\tau_{g,i} \approx 155$ ns computed from the considerations above. Using this method, we obtain $QY = 40 \pm 20\%$. We further note that the NV center can exist in two different charge states ($NV^0$ and $NV^-$) with somewhat different intrinsic lifetimes[S8]. Therefore, the quantum efficiency estimates may be influenced by equilibrium charge population of the NV center, which nevertheless remains predominantly $NV^-$ at excitation wavelengths employed in our experiment[S9].